\documentclass[12pt,preprint]{aastex}

\shorttitle{Densities of Asymmetric Coronal Spectral Profiles}
\shortauthors{}

\newcommand{\kms}{    {$\mathrm{km\,{s}^{-1}}\,$}}
\newcommand{\ang}{    {\AA \,}}
\newcommand{\diagn} {$I_{264}/I_{274}$}

\newcommand{\qcm}  {$\mathrm{{cm}^{-3}}$}

\begin{document}

\title{Core and Wing Densities of Asymmetric Coronal Spectral Profiles: Implications
for the Mass Supply of the Solar Corona}

\author{S. Patsourakos}
\affil{Section of Astrogeophysics, Physics Department, University of
Ioannina, Ioannina GR-45110, Greece}
\email{spatsour@cc.uoi.gr}

\and

\author{J. A. Klimchuk}
\affil{
NASA Goddard Space Flight Center, Solar Physics Lab., Code 671, 8800 Greenbelt Road, Greenbelt, MD 20771, USA
}
\email{james.a.klimchuk@nasa.gov}

\and
\author{P. R. Young}
\affil{College of Science, George Mason University, 4400 University Drive, Fairfax, VA 22030, USA}

\begin{abstract}
Recent solar spectroscopic observations have shown that coronal
spectral lines can exhibit asymmetric profiles, with enhanced
emissions at their blue wings. These asymmetries correspond to
rapidly upflowing plasmas at speeds exceeding $\approx$ 50\kms.
Here, we perform a study of the density of the rapidly upflowing
material and compare it to that of the line core which corresponds
to the bulk of the plasma. For this task we use spectroscopic
observations of several active regions taken by the Extreme
Ultraviolet Imaging Spectrometer of the Hinode mission. The density
sensitive ratio of the Fe XIV lines at 264.78 and 274.20 \AA \, is
used to determine wing and core densities.  We compute the
ratio of the blue wing density to the core density and find that
most values are of order unity.  This is consistent with the
predictions for coronal nanoflares if most of the observed coronal
mass is supplied by chromospheric evaporation driven by the
nanoflares. However, much larger blue wing-to-core density ratios
are predicted if most of the coronal mass is supplied by heated
material ejected with type II spicules. Our measurements do not rule out
a spicule origin for the blue wing emission, but they argue against
spicules being a primary source of the hot plasma in the corona. We
note that only about 40\% of the pixels where line blends could be
safely ignored have blue wing asymmetries in both Fe XIV lines.
Anticipated sub-arcsecond spatial resolution spectroscopic
observations in future missions could shed more light on the origin
of blue, red, and mixed asymmetries.

\end{abstract}


\section{Introduction}

Despite knowing for several decades
that our Sun's surface is enveloped by a multi-million degree
atmosphere, the corona, the mechanism which
supplies it with hot plasmas at several
MK is not currently fully understood.
Whatever the mechanism of coronal mass
supply is, it  should not be considered in isolation
from the source of the elevated
temperatures in the corona. It is rather
that coronal heating
and mass supply represent interlinked
aspects of the conversion of energy
stored in the magnetic fields of the solar atmosphere
into plasma heating and acceleration.
There currently exist two main paradigms
to explain the source of mass into the corona.

The first paradigm, {\it coronal heating and chromospheric
evaporation} postulates that heating takes place in the corona
itself, above the chromosphere \cite[see for example the reviews
by][]{klim06,real10}. Coronal heating could be either steady or more
probably impulsive in nature in the form of small-scale heating
events frequently called nanoflares. These impulsive heating events
drive an enhanced thermal conduction flux and sometimes particle
beams towards the chromosphere leading to heating and evaporation of
chromospheric material into the corona. This plasma eventually fills
in the coronal structures once the impulsive heating is "off". This
paradigm represents the "traditional" view on the coronal mass
supply. Note here, that our use of the term "nanoflare" 
refers to a coronal heating event and
is
generic in that it includes
any impulsive heating event with a small cross-field spatial scale, without regard to physical mechanism.
This can include the dissipation of intense
currents during small-scale reconnection events, as well as of
coronal waves, either generated in the corona or generated in the
chromosphere and then propagating upwards or more possibly some mixture of both.
Impulsive heating events in the chromosphere, 
which are the likely cause of spicules, are not included in this definition.

The second paradigm, type II spicules, places the source of the
coronal mass in the lower solar atmosphere at the chromospheric feet
of coronal structures. Episodic, small-scale expulsions of
chromospheric material into the corona in the form of spicules have
been well-known for a long time \cite[for a recent review on
spicules see][]{tsir12}. The bulk of the "classical" spicules
maintains chromospheric temperatures and returns to the solar
surface. However, recent high spatial and temporal resolution
imaging observations by Hinode and more recently by the Solar
Dynamics Observatory (SDO) showed that there is a second population
of spicules, called type II spicules, which could reach substantial
temperatures of at least 0.08 MK while they shoot up at speeds of
the order of 100 \kms \cite[][]{bdp07,bdp09,bdp11}. Type II
spicules, when  heated during their ascent to even higher,
coronal-like, temperatures could fill in coronal structures with hot
plasma and thus represent the source of coronal mass.

The study of spectral lines represents a powerful tool into the
investigation of the source and the paths of mass exchange in and
between different layers of the solar atmosphere respectively. The
location of the centers of spectral lines, whenever exhibiting
shifts from their rest positions, shows that systematic mass flows
take place. Generally speaking, observations of coronal lines
exhibit blue-shifts while warmer lines formed in the low corona and
the transition region are red-shifted \cite[e.g.,][]{pet99,ter99}. This warm-hot
redshift-blueshift pattern generally complies with the expectations of
impulsive coronal heating, with the blue-shifted material
corresponding to the evaporative upflows and the red-shifted
material corresponding to draining plasma, once the heated material
starts to radiatively cool down and to condense back to the solar
surface \cite[e.g.,][]{carg94}. The magnitude of these bulk motions does not exceed a few
tens of \kms \cite[e.g.,][]{brad08}.

On the other hand, enhancements, even subtle, at the line wings (i.e., away from the almost stationary
line cores) suggest the presence of high speed upflowing, in the
case of blue-wing enhancements, or downflowing, in the case of
red-wing enhancements, plasma within the observational pixels.
As we will see,
 such
emissions, even weak, can have important implications for the origin
and formation of the line cores, which represent the bulk of the
coronal plasma.

Indeed, a series of recent observations
by the Extreme Ultraviolet Imaging Spectrometer (EIS) on Hinode and
the Solar Ultraviolet Measurements of Emitted Radiation (SUMER) on the Solar and Heliospheric Observatory
showed that several
coronal spectral lines exhibit (weak)
excess emissions at their blue wings, or even
secondary components,
at Doppler-shifts larger than $\approx$ 50 \kms,
i.e., beyond  the corresponding line cores
\cite[e.g.,][]{hara08,bdp07,bdp09, mart09,
bryans10,pet10,bdp11,dolla11,mart11,tian11,brooks12,dos12,mc12}. The blue-wing
enhancements are frequently interpreted as
the spectroscopic trace of type II spicules.
The blue-wing enhancements
are not uniformly distributed in space.
For example in active regions the stronger
blue-wing enhancements are found at their edges
\cite[e.g.,][]{hara08,bdp07,bryans10,dos12}.
Note finally that there  are also fewer locations
in the observed fields exhibiting
red-wing enhancements.

In a recent study
\cite{klim12} examined in detail the role of type II spicules in the
upper solar atmosphere and described three separate tests of the
hypothesis that most hot coronal material is supplied by spicules.
He applied two of these tests to published observations (line
asymmetry and ratio of lower transition region to coronal emission
measure) and concluded that the hypothesis must be rejected. In
other words, spicules inject only a small amount of pre-heated
material into the corona (e.g., $< 2\% $ of what is required to
explain active regions). \cite{tri13} subsequently examined whether
type II spicules inject large amounts of material at sub-coronal
temperatures, which is later heated to coronal values at higher
elevations. They concluded that this is not the case, at least not
in active regions and for injection temperatures $\ge 0.6$ MK.

The third test described by \cite{klim12} is based on the
conservation of mass between the fast upflowing spicular material
presumably associated with the observed blue-wing enhancements in
coronal lines and the much slower line core material which makes the
bulk of the coronal emission. This led to predictions of the density
ratio of the two components described above, under the hypothesis
that hot coronal mass is supplied solely by type II spicules. Thus,
it is very timely to perform quantitative measurements of the
densities of the wings of spectral lines with enhanced emission and
to compare with those of line cores. This is exactly the focus of
this study. We hereby determine the density content associated with
enhanced wing emissions of asymmetric coronal spectral profiles and
compare it with that of the line cores. Their ratio represents a
powerful diagnostic of the mass supply mechanism since type II
spicules and impulsive coronal heating events predict largely
different values. To meet these goals we  mainly use observations of
a coronal density sensitive line ratio, and compare the deduced wing
and core densities with analytical as well as simulation-based
predictions for the main candidates of coronal mass supply.

Our paper has the following structure.
In Section 2 we
present the predictions of the density ratio between excess wing and
core emission from various physical mechanisms of coronal mass supply.
In Section 3 we give a brief
recap of our spectroscopic observations; Section 4 supplies a
description of the spectrosopic diagnostic method we use to
determine densities, with a special emphasis given to evaluating and
minimizing the possible impact of line blends on the determination
of the densities of the weak excess wing emissions. Section 5
describes our methodology for inferring the density of enhanced wing
emission and line core from EIS observations, while in Section 6  we compare
them with the theoretical predictions of Section 2. Section 7
finally contains a summary and the conclusions of this study.

\section{Physical Mechanisms of blue-wing Enhancements}
\label{sec:phys} In this Section we deduce theoretical estimates of
the density ratio of the  fast upflowing to core plasma based on
proposed mechanisms for the coronal mass supply discussed in the
Introduction. They are related to type II spicules and coronal
nanoflares.
\subsection{Type II Spicules}
If all coronal plasma comes from heated material at the tips of type
II spicules, then, by conservation of mass,  \citep{klim12}:
\begin{equation}
n_{b}\delta h_{s} = n_{core} h_{core} A,
\end{equation}
where $n_{b}$ and $n_{core}$ are the density of the blue wing upflow
and line core, respectively, $h_{s}$ is the total length of the
spicule, $\delta$ is the fraction of the length that reaches coronal
temperatures, $h_{core}$ is the coronal scale height (the halflength
of the loop strand that contains the spicule), and $A$ is an
expansion factor that accounts for the difference between the
average cross sectional area of the strand and the cross sectional
area where the spicule is observed. $n_{b}$ is simply $n_{diff}$  i.e.,
the density corresponding to the excess blue wing emission as discussed
in more detail in Section 5.
at
locations where there is excess emission in the blue wings of both
Fe XIV profiles (see below).

Following \citet{klim12} we assume typical observed values of:
\begin{eqnarray}
0.05 \le \delta \le 0.3  \nonumber \\
h_{s} =10000 km \, \nonumber \\
A \in [1,3] \nonumber \\
h_{core}= 50000 \, km.
\end{eqnarray}
The range of implied blue wing to line core density ratios is
:
\begin{equation}\label{eq:spic}
%%\left \frac{n_{b}}{n_{core}} \right \vert_{spicule} \in [15,333].
\frac{n_{b}}{n_{core}} \in [16.6, 300] .
\end{equation}
In order to explain the densities observed in the corona,
$n_{core}$, we need a rather large type II spicule density.  For
example, a coronal density of $10^9$ cm$^{-3}$ requires a spicule
density of roughly $10^{11}$ cm$^{-3}$. Such densities have been
measured at chromospheric temperatures in classical spicules
\citep{beckers72,sterling00}; however, $n_{b}$ is the density of the
$\sim2$ MK material at the tips of type II spicules, and those
measurements have never before been made. Two further
comments are now in order. First, the $\delta$ parameter of
Equations and 1 and 2, may be as large as unity, which corresponds
to spicules that are heated to coronal temperatures along their full
length. Larger $\delta$ implies smaller $n_{b}/n_{core}$ ratios (for
example a $\delta$ equal to 1 leads to a minimum density ratio of 5).
However, large $\delta$ is appropriate for only a minority of the
observed type II spicules (Klimchuk 2012). Second, we note that our
calculation of $n_{b}/n_{core}$ is carried out on a per strand
basis, i.e., we work out the anticipated density ratio under the
condition that all the coronal mass of a given strand results from a
type II spicule. Therefore, it does not matter for the calculated
$n_{b}/n_{core}$ values of Equation 3 whether an observational pixel
contains one or more strands (spicules). Finally, if spicules recur
within a given strand at timescales smaller than the time it takes
plasma to drain from the corona, then unobserved temperature
inversions (local temperature maxima) at the base of strands would
be required to explain the corona (Klimchuk 2012).
\subsection{Nanoflares}

The present subsection describes estimates of the  blue-wing
enhancement to line core density ratio based on coronal nanoflares.
In a previous work, we synthesized spectral lines from 1D
hydrodynamic simulations of nanoflares taking place in coronal loops
\citep{pat06}. Depending on the parameters of the nanoflare (e.g.,
energy, etc.) profiles in hot lines ($>5$ MK) can develop
significant blue-wing enhancements, but only if the temperature of
the hot evaporating plasma is "in tune" with the formation
temperature of the line. Much smaller asymmetries are predicted in
warmer coronal lines formed around 1-2 MK. Since the main emphasis
of that paper was given to the (largely) bigger asymmetries found in
the profiles of hot lines that seemed more promising for detecting
signatures of nanoflare heating,  we did not further comment on the
warm lines.

We investigate here in more detail the density content of blue-wing
enhancements as anticipated from nanoflare heating.  In doing this,
we  use both analytical theory and time-dependent 1D hydrodynamic
simulations. We begin with an analytical derivation.
\newline
\textbf{Analytical derivation}
\newline
The evaporation velocity and pressure are greatest at or near the
end of the nanoflare, depending on whether the heating profile is
square-like or triangular-like.  Velocity decreases quickly
thereafter, because the temperature and heat flux decrease rapidly
and because density increases \citep[e.g.,][Fig. 2]{klim06}.
Pressure decreases more slowly because the temperature decrease is
largely offset by the density increase (only radiation causes the
pressure to decrease, and the radiative cooling time is long).

We are interested in the density of the upflow as measured in the
blue wings of lines formed at $T \simeq $ 2 MK.  Most of this emission
comes from the transition region during a time interval immediately
following the nanoflare, when density is still increasing.  We
ignore the flash of emission that occurs as coronal plasma rapidly
heats through this temperature early in the nanoflare for the
following reasons:  it is very temporary, the density is still low,
and the velocities are relatively slow.

Because pressure decreases only modestly during the conductive cooling (evaporation) phase,
the pressure at the end of this phase (subscript “*”) is approximately equal to the pressure
at the time of the blue wing upflow (subscript “b”) \citep{carg12a}:
\begin{equation}
P_{*} \approx P_{b}.
\end{equation}
Thus,
\begin{equation}
n_{b} \approx n_{*}(T_{*}/T) .
\end{equation}

The emission in the line core is produced primarily by the coronal
plasma as it cools, relatively slowly, through $T$.  For lines that
are not too hot, this is during the radiative cooling phase that
follows the conductive phase.  According to \citet{carg04}, there is
a $T \propto n^{2}$ scaling during this phase, so the downflow
density is
\begin{equation}
n_{c} = {(T/T_{*})}^{1/2}n_{*}.
\end{equation}
Combining the above two equations we get:
\begin{equation}
n_{b}/n_{c} = {(T_{*}/T)}^{3/2}
= {(T_{*}/T_{m})}^{3/2} {(T_{m}/T)}^{3/2},
\end{equation}
where $T_{m}$ is the maximum coronal temperature,
at the end of the nanoflare.
According to equation (A1) of \citet{carg04},
\begin{equation}
T_{*}/T_{m} = {({\tau}_{c}/{\tau}_{r})}^{1/6},
\end{equation}
where ${\tau}_{c}$ and ${\tau}_{r}$ are the conductive and radiative
cooling times at the end of the nanoflare. The cooling time ratio is
typically in the range ${10}^{-4} <{\tau}_{c}/{\tau}_{r} <
{10}^{-2}$ at the end of the nanoflare \citep{klim08,carg12b}, so
\begin{equation}
0.22 < T_{*}/T_{m} < 0.46.
\end{equation}

For $T$=2 MK and $T_{m} \in [5,15]$ MK, we have an enhanced
blue-wing emission to core density ratio of
\begin{equation}\label{eq:nano1}
\frac{n_{b}}{n_{core}}   \in [0.4, 6.4]
\end{equation}

\textbf{1D Hydrodynamic Nanoflare Simulations}
\newline
The analytical prediction of Equation \ref{eq:nano1} is confirmed by
1D hydrodynamic nanoflare simulations. For this task we use the 1D
hydrodynamic code called ARGOS, which solves the time-dependent
hydrodynamic equations on a adaptively refined grid \citep{anti99}.
We consider semi-circular loops with lengths of 50 and 100 Mm. The
selected lengths correspond to  typical loop sizes  in active region
cores and peripheries respectively. The loops are  assumed to lie
vertically above the solar surface. Initial conditions are
determined by calculating static equilibrium solutions with  peak
temperatures of $\approx$ 0.3 MK and 1 MK. Each initial condition is
submitted to a heating pulse with a duration of 50 s and is then
allowed to cool down for another 5000 s. We consider heating pulses
with different magnitudes leading to maximum temperatures in the
range [3.5,17] MK. The impulsive heating for each pulse is uniformly
distributed along the modeled loop. The solution of the
time-dependent hydrodynamic equations supplies the plasma
temperature, density and bulk flow speed as functions of location
along the loop and of time. Knowing these physical parameters we
produce synthetic profiles for the Fe XIV 264 and 274 lines using
CHIANTI.  Each profile is given a thermal width from the local
temperature and is Doppler-shifted by an amount given by the
line-of-sight projection of the local bulk speed. All profiles are
calculated in wavelength grids with the same spectral resolution as
the blown-up profiles we will describe in Section \ref{sec:anal},
i.e., 50 times the EIS spectral resolution. In the calculation of
the Doppler-shifts, we assume that the modeled loops are viewed from
above.  The instantaneous profiles are then averaged both in time
and space in order to emulate spectroscopic observations of a
multi-stranded coronal loop heated by nanoflares. We produce two
types of average profiles meant to emulate spectroscopic
observations from the footpoint and upper (coronal) parts of a
coronal loop. The {\it footpoint} profile combines the profiles in
the interval spanning from  the deepest point in the transition
region during each simulation  and 10 Mm upwards. Since the location
of the transition region is changing in response to pressure changes
in the corona during impulsively heated events, selecting the
deepest point of the transition region during the entire simulation
ensures that the transition region is covered at {\it all} times. A
{\it coronal} profile is constructed from all profiles above the
computation region of the footpoint profile until the loop apex.
Both footpoint and coronal profiles were averaged over the entire
duration of the corresponding simulation. In addition, a "total"
average profile over the entire loop (footpoints and coronal
section), is also produced, in order to emulate an unresolved
observation combining both footpoint and coronal sections.

We then apply the procedures we describe in
Section 5
for the observational data to determine
the ratio $R$ of the blue-wing enhancement
to the core density from the synthetic
profiles resulting  from our nanoflare simulations
for the footpoint, coronal and "total" profiles.
The applicability of standard spectroscopic diagnostics, like
the density-sensitive line ratios used here,
to multi-thermal plasmas expected from nanoflare heating
was demonstrated by \citet{klim01}.
The resulting range of the blue wing to core density ratio
is:
\begin{equation}\label{eq:nano2}
\frac{n_{b}}{n_{core}} \in [0.44, 1.01]
\end{equation}
and lies within the interval predicted by the analytical derivation of
Equation \ref{eq:nano1}.

Note here that our results above, seem  to be  insensitive to
the spatial localization of  nanoflare heating. To check this we perform
a set of additional 8 nanoflare simulations with heating concentrated
towards the strand footpoints. We consider a 50 Mm long
strand at an  initial (static equilibrium) temperature of 1 MK.
For each simulation, the strand is submitted to a single heating pulse,
with an exponential drop-off above
the initial location of the transition region.
$1/e$ heating lengthscales of 1000 and 5000 km are used.
The simulated nanoflares
reach maximum temperature in the range 3-12 MK.
The $n_{b}/n_{core}$ ratio  takes values in the interval 0.55-1.2, which
is similar to the range obtained for uniform heating.

Before proceeding to compare our theoretical predictions with
observations, we note the blue wing to core density ratios predicted
for type II spicules (Equation \ref{eq:spic}) and coronal nanoflares
(Equations \ref{eq:nano1} and \ref{eq:nano2}) are greatly different.
The ranges do not overlap.
Consider the upward flow of hot material crossing an imaginary surface in the low corona.  The time-integrated mass flux must equal all of the coronal material that ultimately fills the strand.  In the case of a spicule, the material crosses the surface very quickly.  The crossing time scale is just the thickness of the hot spicule tip divided by the upflow velocity, which is about 10 s.  Evaporation, on the other hand, is a process that continues for a much longer time.  Since the velocity is similar (we measure a particular wavelength band in the blue wing), the density must be inversely proportional to the time scale in order to get the same total mass.  Hence, the density is much smaller for evaporation than for spicules.
It is unreasonable to expect that the
uncertainties and limitations in the determinations of the density
ratios would bridge the significant gap. This makes the upflow to
core density ratio an important  and sensitive diagnostic means for
studying the source of high-speed coronal upflows.

\section{Observations}
We use spectroscopic observations
taken by the Extreme Ultraviolet Imaging Spectrometer
(EIS), \cite{culh07}, on-board the Hinode mission.
EIS is a normal incidence spectrometer taking
observations in two wavelength ranges in the Extreme Ultraviolet:
170-210 \AA\, and 250-290 \AA. These ranges cover
lines formed in the chromosphere, transition region
and corona. EIS has spectral pixels of 22 m\AA, corresponding to $\approx$
25 \kms and a spatial resolution of about 2 arcsec. For more
on EIS we refer to  \citet{culh07}.

The observations we analyze here were taken during the period
2006-2007 and correspond to several two-dimensional spectral scans
(rasters) over active regions at various locations on the solar
disk. Detailed spectra of various selected spectral lines at each
spatial location covered by the raster scan are obtained. More
information (e.g., time of  observation, size of raster, location of
raster etc) for each observation is supplied in Table-1. The  level0
data were submitted to the standard data reduction pipeline for EIS
data using the \verb#eis_prep.pro# routine.

\section{The Fe XIV (264/274) Density Sensitive Ratio}
\label{sec:diagn} The intensity ratio of the  Fe XIV lines at 264.78
and 274.20, hereafter 264 and 274, represents a good density
diagnostic for coronal plasmas. All atomic physics calculations and
diagnostics used in this paper use the latest version  (v7.1,
\cite{lan13}) of CHIANTI \citep{dere97}.

\begin{figure}
\epsscale{1.00} \plotone{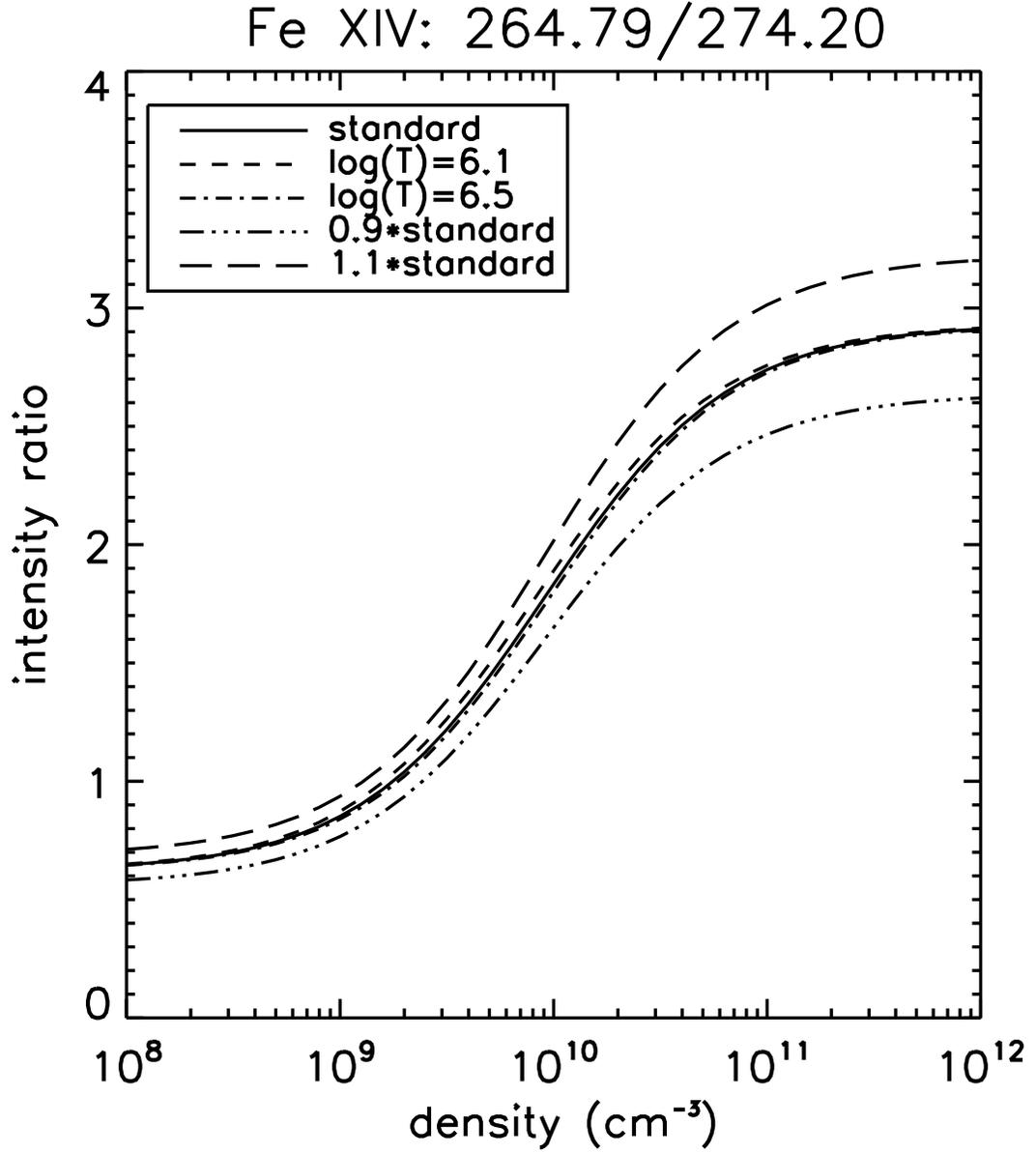}
\caption{Fe XIV
264/274  density diagnostic. Solid line: standard curve used in all
calculations for the formation temperature of Fe XIV (6.3 in
log(T)); short-dashes: log(T)=6.1; dashes-dots: log(T)=6.5;
dashed-triple-dots: 0.9*standard curve; long-dashes:1.1*standard
curve.} \label{fig:diagn}
\end{figure}

The solid line of  Figure \ref{fig:diagn} shows the theoretical 264
to 274 intensity ratio against density for the formation temperature
of Fe XIV (6.3 in log(T)). This  curve is used throughout this study
for all density calculations and we call it our standard diagnostic
curve. The 264/274 intensity ratio is  sensitive to density for the
interval  $\approx$ [0.65,2.9] which corresponds to densities in the
range $\approx {10}^{8}-{10}^{12} \mathrm{{cm}^{-3}} $. However, the
region of higher sensitivity corresponds to ratios in the more
compact ratio range of $\approx$ [0.9,2.6] and conversely to densities in
the range of $\approx [2\times{10}^{9}, 4\times{10}^{10}]
\mathrm{{cm}^{-3}} $. For intensity ratios approaching   the lower
or the  upper limit of the former interval, where the density
sensitivity on the line ratio becomes much smaller (i.e., the
density-ratio curve becomes almost flat), one can only deduce an
{\it upper}, $n_{max}$ =${10}^{8}$ \qcm  or {\it lower}
$n_{min}={10}^{12}$ \qcm \, limit on the density respectively.
Note here  that one should not over-intepret these densities
and their values;
cases with densities equal to   $n_{max}$ or $n_{min}$
are  simply telling us that the corresponding
intensity ratio measurements are below/above the low/high limits
respectively.

A potential burden in the analysis of the weak wing emissions of
spectral lines, which is the focal point of this study, could arise
from the presence of line blends, i.e., "parasitic" lines close to
the "target" lines. Line blends may affect the intensities of the
"target" lines. Inspection of published EIS line lists, based on
both atomic physics calculations and observations shows that the two Fe XIV lines of interest are  largely free of
blends  \cite[e.g., ][]{youn07,brown08,gdz12}. However, there exist
two lines close to the centers of
the two Fe XIV lines that warrant meticulous consideration. First,
there is a Si VII line at 274.18\ang, or $\approx$ 22 \kms towards
the blue of the 274 Fe XIV rest wavelength. Second, there is  a Fe
XI line at 264.77\ang, or $\approx$ 18 \kms towards the blue of the
264 Fe XIV rest wavelength. These lines are not very far from the
cores of the Fe XIV lines of interest, but they may nevertheless
affect their wing intensities. For example \citep{land09} found
that the Si VII line contributes  57$\%$
to the observed fan loops in an AR.

Fortunately, the  Si VII and Fe XI blends have known intensity
ratios with a Si VII line at 275.35\AA\, and a Fe XI line at 188.23\AA\,
respectively \citep[e.g.,][]{youn07}. These two lines  can be
observed by EIS. Moreover they  are isolated and thus their
intensities can be measured accurately. For typical coronal
densities the Fe XI  264.77/188.23 ratio varies between about 0.026
to 0.043; therefore measuring the Fe XI 188.23 \ang intensities and
multiplying them by  0.043 supplies an {\it upper} limit on the
contribution of the Fe XI 264.77 line to the Fe XIV 264 line.
Similarly, the Si VII 274.2/275.35 intensity ratio has a maximum
value of 0.25; therefore multiplying the Si VII 275.35 intensity by
0.25 supplies an {\it upper} limit on the contribution of the Si VII
274.2 line to the Fe XIV 274 line.

Observed cases for which the Fe XI and Si VII blends are
simultaneously $\le 10 \%$  of the measured intensities of the Fe
XIV 264 and 274  lines respectively were selected for further
analysis since the blend contribution to the "target" line core and
wing intensities can  be rather safely neglected. As discussed
before, both line blends are only slightly offset and roughly at the
same distance
 from the rest wavelength positions of the "target" lines.
Therefore, given our selection criterion, they should amount at
maximum to no more than the $10 \%$ of the "target" line  core.
The blends will contribute an even
smaller percentage to the line wings if the flows that produce the
wing emission have a temperature dependence such that cooler plasma
is also slower. This will be the case for evaporated plasma, where
$v \propto T$, though not necessarily for spicules. The cases where
the $10 \%$ criterion is satisfied for both Fe XIV lines are
referred to as ``No-Blend'' (second column in Table
\ref{table:gstats}). On average, the blend contribution for the
cases not satisfying the above criterion for the "base" observation
discussed later on in the paper correspond to 16 $\%$ and  19 $\%$
of Fe XIV 264 and 274 intensities respectively. The average
intensity contributions of Fe XI and Si VII to the 264 and 274
features across all the active region raster of the "base"
observation are 7 $\%$ and 4$\%$, respectively.

We now consider the potential impact of our "No-Blend" criterion on
the density diagnostic. First consider the case when the Fe XI and
SI VII blends amount to exactly the same fraction of the 264 and 274
lines respectively, then they will not influence the density
diagnostic at all: i.e. the standard diagnostic curve of Figure
\ref{fig:diagn} will not be affected. Allowing for a blend
contribution of a  maximum value of 10 \% in either the 264 or the
274 intensity will lead  to an  overestimate or underestimate of the
264/274 intensity ratio by a factor of 0.1 respectively. In Figure
\ref{fig:diagn} we show diagnostic curves corresponding to these two
extreme cases: long-dashed and dashed-triple-dotted corresponding to
1.1 and 0.9 of the standard diagnostic curve. When the 264 and 274
blends represent a smaller, yet non-equal amount of these lines,
their contribution to the density diagnostic will be bounded between
the two extreme cases considered above: i.e., the corresponding
diagnostic curves will be between the long-dashes and
dashes-triple-dots of Figure \ref{fig:diagn}.

We can now evaluate the impact of the blend on the density
diagnostic. Take an intensity ratio of 1.1. The standard curve for
this ratio yields a typical AR density of about 2$\times {10}^{9}$
\qcm. Allowing for a maximal blend contribution of 10 \% of either
the 264 or the 274 line (i.e., the long-dashed and  the
dashed-double-dotted curves of Figure \ref{fig:diagn} respectively)
leads to an overestimate or underestimate of the density by a
factor of less $\approx$ 2 (we use the standard curve in all density
calculations). Similar discrepancy factors can be found if we
consider other values for the line ratio. Note that these
underestimate/overestimate factors represent generous {\it upper}
limits.

As discussed before the  standard density diagnostic curve of Figure
\ref{fig:diagn} was determined using the temperature of peak Fe XIV
formation under equilibrium ionization conditions (log(T)=6.3).
However, it is possible that the plasma is at a different
equilibrium temperature or that the plasma is rapidly heating as it
rises or rapidly cooling as it descends. Therefore, the actual
temperature at the time Fe XIV is present could be different from
the equilibrium temperature.   We thus calculated the theoretical
264 to 274 intensity ratio for log(T)=6.5 and log(T)=6.1 long-dashes
and dashed-dots of  Figure \ref{fig:diagn} respectively. These
temperature correspond to the upper(lower) end of the $FWHM$ range
of the Fe XIV contribution function. From  Figure \ref{fig:diagn} we
have that a hotter upflowing plasma leads to an overestimate of the
density with respect to the standard density diagnostic curve, but
by a rather small factor (much less than 2). The opposite
(underestimating the density) occurs when considering warmer,
downflowing plasmas (dashed curve in Figure \ref{fig:diagn}). Again,
the deviation from the standard diagnostic curve is rather small.

To sum up,
we conclude that the possible contribution of (known) line blends
and departures from the formation temperature
of Fe XIV leads to density  uncertainties
of a factor of 2.

Before proceeding, we note that other coronal density
sensitive line ratios available in EIS observations, Fe XII
196.64/195.12  and Fe XIII 203.82/202.04, exhibit also  blending issues
or have other strong lines in their vicinities that need to be
treated simultaneously \cite[e.g., ][]{youn07}.

\section{Analysis Method}
\label{sec:anal}

We are now set to describe the method we use to calculate the
density associated with excess blue or red wing emissions which
correspond to fast upflowing or downflowing plasmas respectively.
These densities will then be compared with those of the line core
corresponding to almost static plasmas. To increase the
signal-to-noise (S/N) ratio, we calculated  averaged line profiles
over 3$\times$3 full resolution pixels. It is important to have
sizeable S/N ratios since we wish to obtain detailed measurements at
the line wings, where intensities are low.

For all macropixels satisfying the "No-Blend"
criterion of the previous Section we proceed as follows:
\begin{itemize}
\item interpolate  each 264 and 274 Fe XIV profile
and the associated intensity errors (arising from photon-counting
statistics and readout noise) onto a 50 times finer grid using an
improved cubic-spline interpolation scheme. The scheme, fully
described in Klimchuk, Patsourakos and Tripathi (2013), addresses the important fact that while  the intensity
we observe in a spectral bin is the mean intensity averaged over the
bin, traditionally it is assigned to the bin center. This is only
appropriate, however, if the line profile within the bin is
symmetric about the bin center (e.g., a straight line segment),
which is generally not the case. The true bin center intensity is
generally different from the bin average, and not taking this into
account can impact the very delicate measurements of wing
intensities and line center locations we hereby perform. The new
scheme, based on an iterative method, rapidly converges to solutions
conserving the total intensity across the bin. It is therefore
called Intensity Conserving Spline Interpolation (ICSI).
\item
determine the
wavelength location of  peak intensity
which is assumed to correspond to a zero Doppler-shift;
\item determine the integrated intensity of
the blown-up (50 times higher resolution) profiles for their blue
([-150,-50] \kms from peak intensity location) and red ([50,150]
\kms from peak intensity location) wings: $i_{blue}$ and $i_{red}$
respectively;
 \item determine the residual wing intensity
$i_{diff}=i_{blue}-i_{red}$
for both Fe XIV lines;
\item if  $i_{diff}$ has the same sign for {\it both} lines, i.e.
both Fe XIV lines exhibit a blue-wing or red-wing asymmetry, then use the corresponding $i_{diff}$
to determine
the corresponding density $n_{diff}$ from the \diagn diagnostic of Section 4;
\item subtract a linear background from the blown-up profiles
and determine the  core
intensity $i_{core}$ in the [-30,30] \kms interval centered
at the peak intensity location,
and the associated density $n_{core}$ from the \diagn diagnostic of Section 4;
\item  deduce a  blue-red
asymmetry, $BR = i_{diff}$,
and its normalized to the core intensity version,
$BR_{norm}=i_{diff}/i_{core}$;
\item determine the density ratio, $rat=n_{diff}/n_{core}$
between  excess wing  and core emissions.
\end{itemize}

Profiles with positive $BR$ correspond to a  blue-wing enhancement
while those with negative  $BR$  to a red-wing enhancement. Similar
analysis methods to deduce profile asymmetries are commonly used
\cite[e.g.][]{bdp07}. Note here that a significant number of
profiles corresponds to residual intensities, $i_{diff}$ with
$264/274$ ratios lying outside the density sensitive part of the
$n-264/274$ curve  discussed in Section \ref{sec:diagn}. For these
profiles we are  only able to deduce an upper (${10}^{8}  \mathrm{
{cm}^{-3}} $) or a lower (${10}^{12} \mathrm{{cm}^{-3} }$) density
limit.

An underlying assumption in our analysis is that, 
by differencing the two sides of the line profile, we are isolating the emission of 
rapidly flowing plasma from the wing emission of rest material. This is strictly correct only 
if fast material contributes to just one side of the profile, i.e., is only flowing up or down. 
The assumption is nonetheless reasonable if there are both upflows and downflows as long as the enhanced emission 
is substantially brighter in one wing than the other. We find that $n_{diff}$ is not well correlated with either the magnitude 
or sign of ${BR}_{norm}$, suggesting that measurements 
are a good indication of the actual density of the dominant high-speed material.

Note here that in the calculation of the $n_{diff}/n_{core}$ ratio
any systematic error, e.g. instrument calibration and atomic
physics, cancels out since it  has the  same  sign for each
calculated density. Moreover, and since we are interested in a
density ratio, it may be that the impact of the line blends of the
264 and 274 lines discussed in Section \ref{sec:diagn} could cancel
out as well. This is because, as discussed in Section
\ref{sec:diagn} both blends are at similar wavelength locations with
respect to the line of interest and therefore it is possible they
may affect the wing and core intensities and thus densities in a
similar manner. At any rate, we  need to consider the maximum
uncertainty factor of $\approx$ 2 in the inferred densities from
(known) line blending we deduced in Section \ref{sec:diagn}.

We therefore need to consider only the effect of random errors in
$n_{diff}/n_{core}$ ratio. Uncertainties in the resulting
histograms are deduced with the
following bootstrapping method.
For a number of realizations (=1000) we randomly perturb
the density ratios for a random sample of $N$macropixels, where
$N$ is the  total number of macropixels
\footnote{Note,
that a given macropixel may be chosen 0,1, or several times.}
 and determine the corresponding histogram. We then determine the average and standard
deviations of the resulting histograms for the 1000 realizations.
The pertubations in the density ratios take into account standard errors  and error propagation.

\section{Results}
\label{sec:results}

\begin{figure}
\epsscale{1.00} \plotone{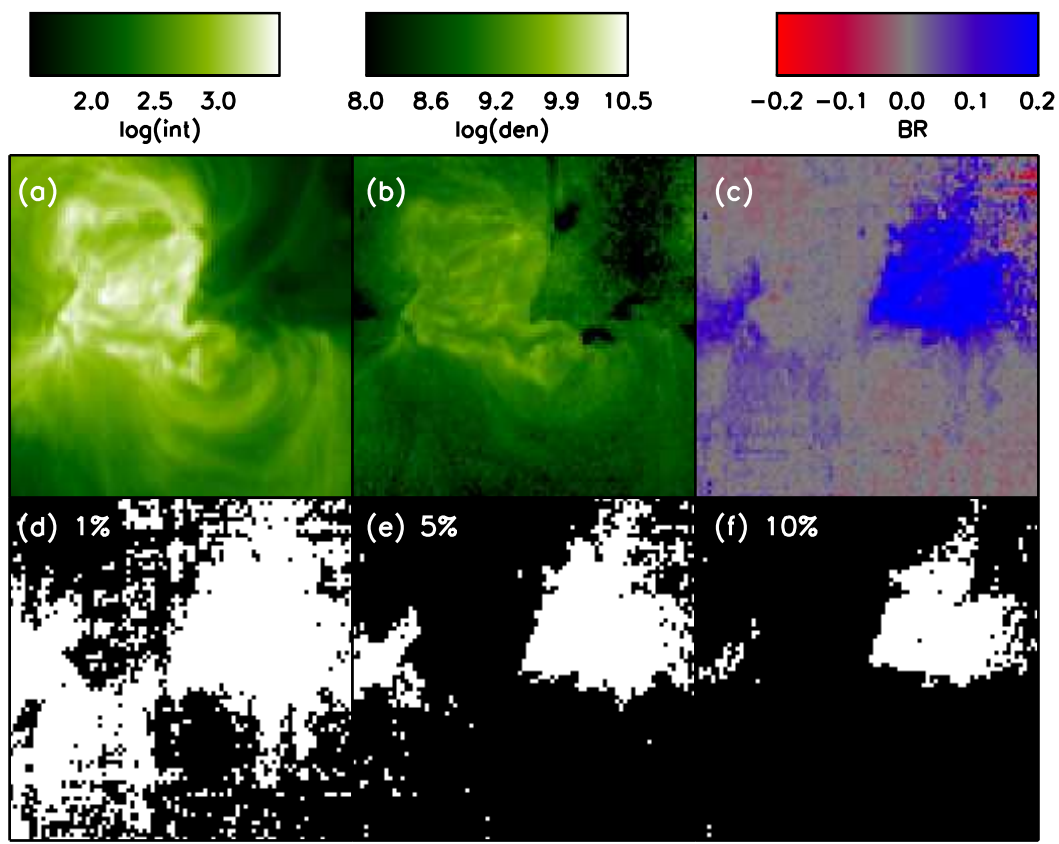}
\caption{Observations of
11 December 2007 (dataset 8 of Table-1). Panel (a): intensity image
for Fe XIV 274 (log scaling). Panel (b): density map from the
intensities integrated over the 264 and 274 profiles (log scaling).
Panel (c): normalized blue-red asymmetry (c.f. Section
\ref{sec:anal}; linear scaling- positive values (blue) correspond to
blue-wing asymmetries; negative values (red) correspond to red-wing
asymmetries. Panels (d),(f),(g): binary masks with locations
having ${BR}_{norm} >$ 0.01, 0.05 and 0.1 respectively displayed
in white.
Observed field of view is 256$\times$256
$\mathrm{{arcsec}^{2}}$.} 
\label{fig:imgs}
\end{figure}

We now present the results of our analysis. Take as example the
observation which took place on 11 December 2007 (dataset 8 of Table
1); we call this observation our base observation. We will
nevertheless supply all the important information regarding all
considered observations. In Figure \ref{fig:imgs} we show the 274 Fe
XIV image (panel a), the density map from the profile-integrated
264 and 274 intensities using the density diagnostic of Section
\ref{sec:diagn} (panel b), and the normalized $BR$ asymmetry as
defined in the previous Section (panel c). From the 274 image we
note we are dealing with  a small bipolar  AR. Densities are
enhanced in the AR core, and they decrease when going to its
periphery. Most of the observed field contains profiles with $BR>0$,
in other words with enhanced blue-wing emission, in agreement with
previous studies. The blue-wing asymmetries are stronger at the
edges of the AR (dark areas in the 274 intensity image), although
they can be also seen in the AR core. There are finally fewer
locations with red-wing enhancements ($BR<0$).
Note here, and also found in other studies,
that the observed asymmetries are rather weak. For example, for profiles
with $BR>0$, the average $BR_{norm}$ is only 0.04; 76$\%$
of the profiles have $BR_{norm} > 0.01 $.
To illustrate this furthermore, panels (e),(f) and (g) in Figure
2 contain binary masks with the locations, shown in white, of 274 profiles 
having $BR_{norm}>$  0.01, 0.05  and 0.1 respectively. It is obvious
that locations with sizeable $BR_{norm}$ (i.e., $> 0.05$ for example)
are signficantly fewer than those corresponding to the full $BR_{norm}$ 
distribution.

In the left panel of  Figure \ref{fig:distr2} we plot the histogram
of $n_{diff}/n_{core}$ for points with $BR>0$ of the base
observation as determined by the procedure described in Section
\ref{sec:anal}. In this case $n_{diff}$ is equal to $n_{b}$, the
density of the blue-wing enhancement.  The peak of the distribution
is below 1 ($\approx$ 0.2) signifying that the most probable density
of the blue-wing enhancement is smaller than that of the line core.
The median value of the distribution is slightly above 1 (2.2).
These findings are consistent with the predictions for coronal
nanoflares using both analytical (Equation \ref{eq:nano1}) and
numerical (Equation \ref{eq:nano2}) approaches.  On the other hand,
the density ratios are much smaller than predicted for type II
spicules.  If spicules supply most of the hot plasma in the corona,
then $n_{diff}/n_{core}$ must be significantly larger than 1 ($>>$
16.6): see Equation \ref{eq:spic}).  It may be that the observed blue wing
emission does in fact come from spicules and has a density $n_{diff}
= n_b$, but the density of this material will be far less than
$n_{core}$ after it has expanded to fill the loop strand. In that
case, the observed $n_{core}$ represents material not associated
with spicules (e.g., evaporated material produced by coronal
nanoflares).  We note that a small minority of density ratios at the
tail of the distribution, having large
($>>1$) $n_{diff}/n_{core}$,
are
consistent with type II spicules being the dominant source
of hot plasma at those locations.

The two leftmost panels of Figure \ref{fig:distr3} show results for
macropixels with $BR>0$ in the base observation where only an upper
limit (first panel) or a lower limit (second panel) could be
determined for the blue-wing enhancement, as discussed in Section
\ref{sec:diagn}. Cases with blue-wing enhancement density equal to
an upper limit $n_{max}$ =${10}^{8}$ \qcm are not consistent with
type II spicules since this density is smaller than the
corresponding core density (1st panel of Figure \ref{fig:distr3}).
On the other hand, cases with blue-wing enhancement density equal to
a lower limit $n_{min}={10}^{12}$ \qcm are  consistent with type II
spicules since they correspond to blue-wing enhancement to core
density ratios much higher than 1 (2nd panel of Figure
\ref{fig:distr3}). Interestingly, EIS observations of coronal jets
show elevated blue wing  densities in the range found above
\citep{chi08}. We defer a discussion of results from macropixels
with negative $BR$ asymmetry displayed in Figures \ref{fig:distr2}
and \ref{fig:distr3} for the next Section.

In order to relate how the densities associated with excess wing
emissions depend on different solar features within the observed
ARs, we display in Figure \ref{fig:fan} the variation of
$n_{diff}/n_{core}$ as a function of $n_{core}$ for locations
exhibiting $BR>0$ (upper panel) and $BR<0$ (lower panel). The plot
corresponds to the base observation. Locations with low $n_{core}$
correspond to low-emitting structures that can be found in AR edges,
while locations with high  $n_{core}$ correspond to AR cores. From
Figure \ref{fig:fan} we note a trend of decreasing
$n_{diff}/n_{core}$ with increasing $n_{core}$ for locations with
both $BR<0$ and $BR>0$.

As a further illustration of the spatial distribution of the
density ratio, Figure \ref{fig:spat} contains $n_{diff}/n_{core}$
for profiles with $BR>0$ for different ranges. We can note that
profiles with  high  $n_{diff}/n_{core}$ (green crosses), which are
consistent with spicules, are located mostly at the observed AR
edges, although there are also a few cases in the AR core. The more
numerous profiles, with low $n_{diff}/n_{core}$ values (red
crosses), which are consistent with nanoflares, occur both in the AR
core and at its edges. A similar behavior applies to profiles with
intermediate $n_{diff}/n_{core}$ values (blue crosses).

Having available the densities of the excess wing emissions and of
the line cores we are able to calculate the formation path lengths
of the corresponding emissions. We first note that the
intensity of neither 264 nor 274 has a $n^{2}$ dependence: the
intensities of both lines are $\propto$ $n^{2+\alpha}$ with $\alpha
>0$ for 274 and $\alpha < 0$ for 264. It is indeed because of these
dependencies that the intensity ratio of 264 and 274 is
density-sensitive. However, the {\it sum} of the  264 and 274
intensities does almost have an exact $n^{2}$ dependence. We can
therefore write that:
\begin{equation}\label{eq:ff}
I_{264}+I_{274}=F(T) n^{2} \phi dl,
\end{equation}
where $I_{264}$ and $I_{274}$
correspond to the 264 and 274 intensities respectively of a
particular spectral feature (excess wing emission or core),
$F(T)$ is the sum of the contribution functions of 264 and 274
calculated at the formation temperature of Fe XIV multiplied by the
iron abundance, and $dl$ is the line-of-sight thickness of the volume containing emitting material, and $\phi$ is the filling factor of that material.  For blue wing observations near disk center, $dl$ is 
either the thickness of the transition region
(for a nanoflare)
or the length of the hot section of the spicule, and $\phi$ is the fractional area of the unresolved strands that are experiencing either nanoflare evaporation or spicule ejection.  For line core observations, $dl$ is the scale height of the emission (thickness of the corona) and $\phi$ is fractional volume of the strands that contain plasma at that temperature (2 MK in this case).

We applied Equation \ref{eq:ff}
to the line
core and excess wing intensities and densities
to deduce $\phi dl$. Figure \ref{fig:ff}
contains the results for
the core (upper panel) and the
excess wing emission (lower panel)
for $BR>0$ for the base observation.
We note that the excess wing emissions have
significantly smaller $\phi dl$
compared to the line cores, with  the corresponding
distributions peaking at $\approx$
300 km and 4000 km respectively.
The latter is comparable to diameters
of  macroscopic coronal loops  \cite[e.g.,][]{pet13}; even longer $\phi dl$
of the line cores could correspond
to diffuse corona regions, with very long
lines of sight. The much smaller formation
lengths of the excess blue-wing emissions hint
on small-scale possibly time-dependent processes.
Both type II spicules
 \cite[e.g.,][]{bdp09}
and nanoflares involve
small spatial scales.
\newpage

\section{Summary and Discussion}

\begin{figure}
\centerline{\hspace*{0.015\textwidth}
               \includegraphics[width=0.49\textwidth,clip=]{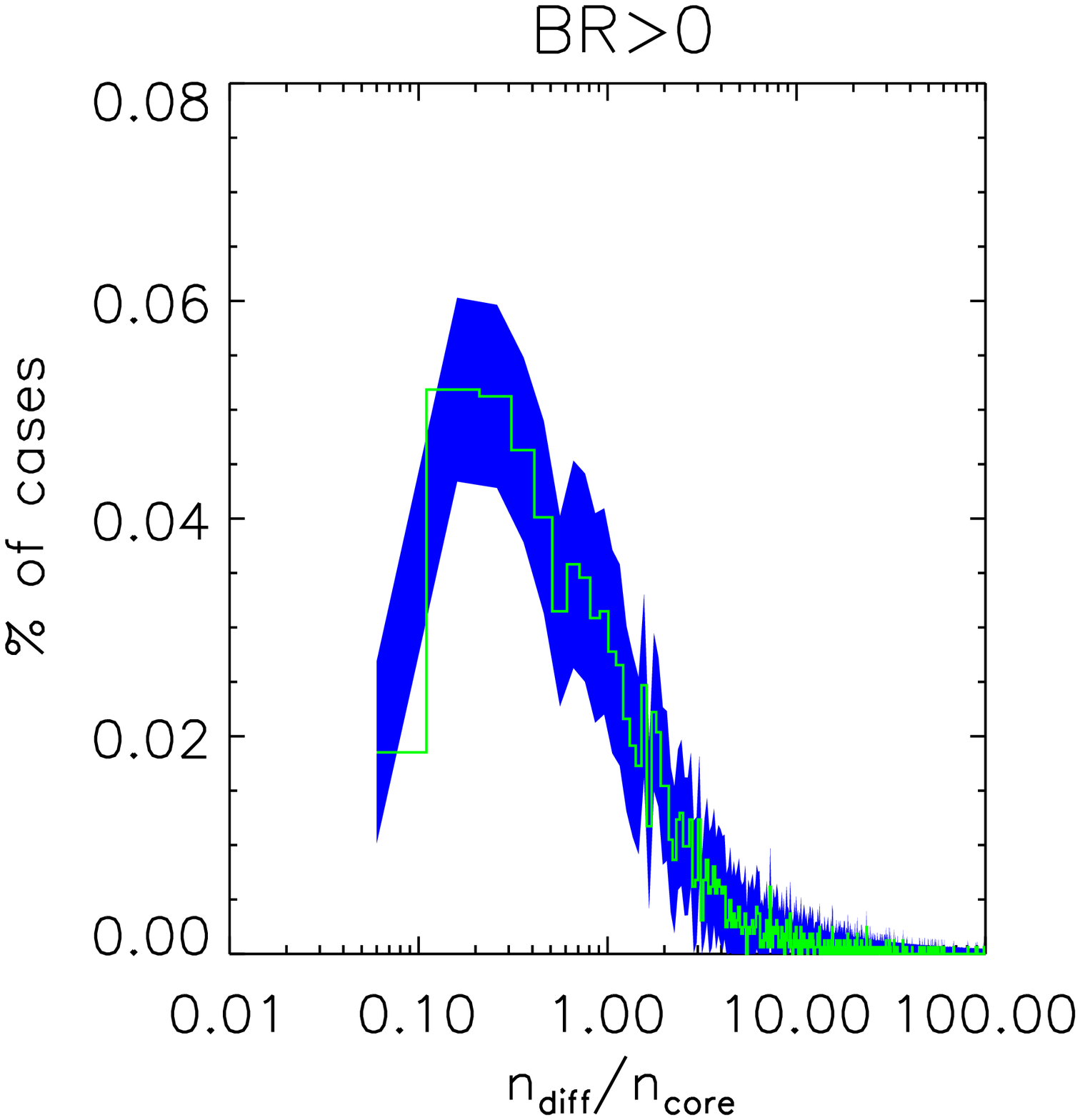}
\hspace*{-0.01\textwidth}
               \includegraphics[width=0.49\textwidth,clip=]{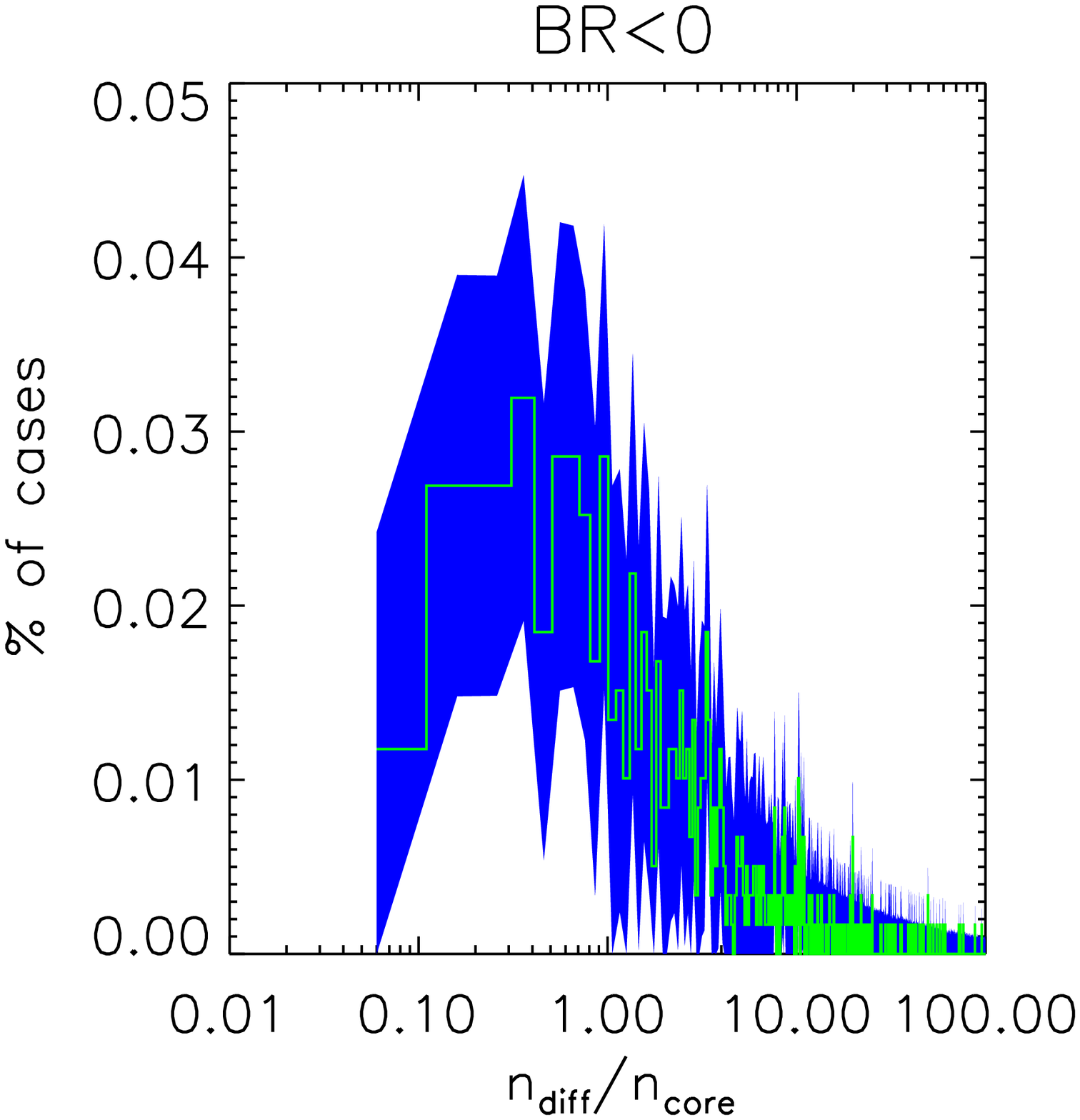}
}
\caption{Histograms (green histogram-mode lines)
and 2-$\sigma$ uncertainties (blue contours)
of $n_{diff}/n_{core}$ for
the observation of 11 December 2007. Locations with positive blue-red
(left panel) and negative blue-red asymmetry (right panel).}
\label{fig:distr2}
\end{figure}

\begin{figure}
\epsscale{1.00}
 \centerline{\hspace*{0.015\textwidth} 
               \includegraphics[width=0.49\textwidth,clip=]{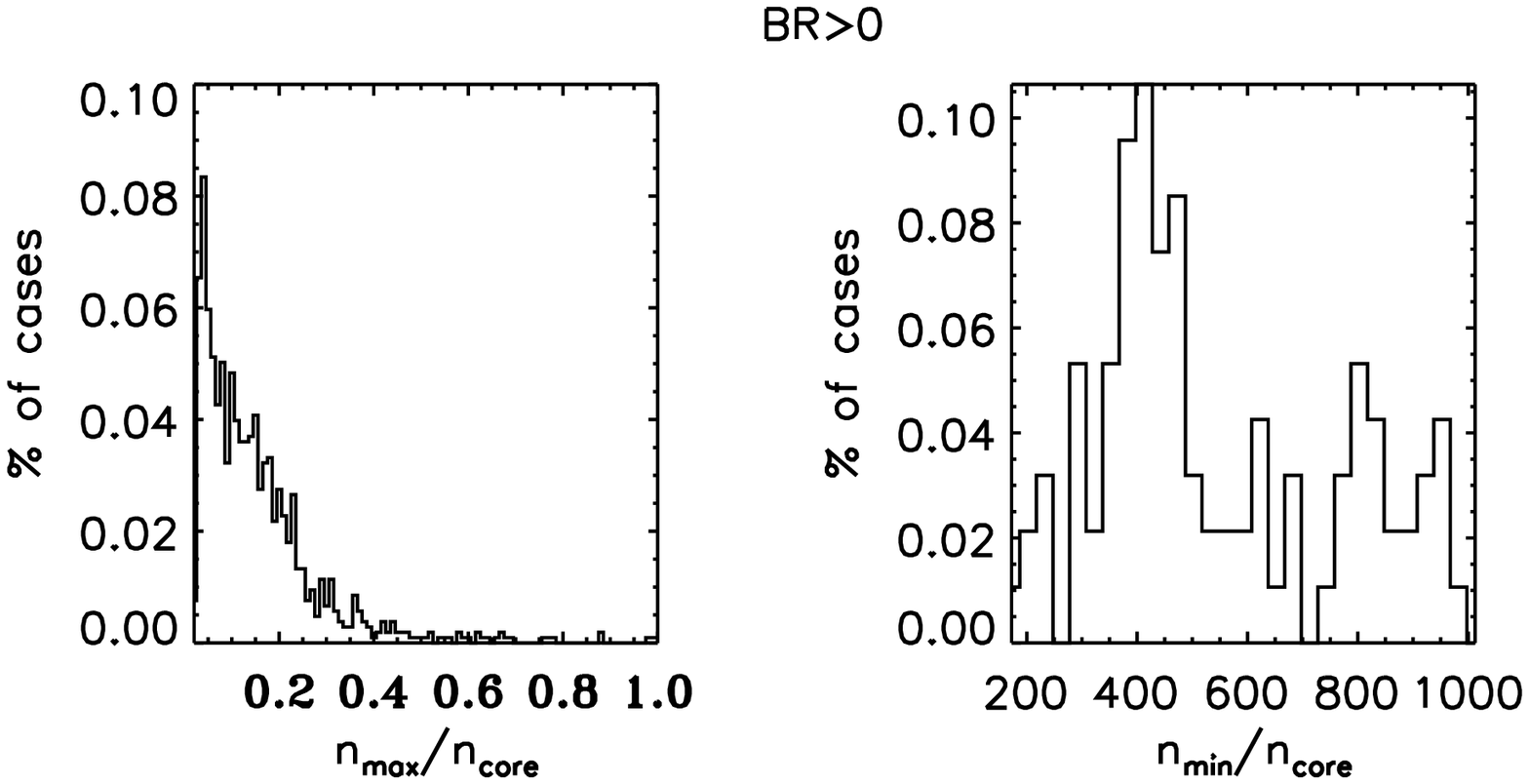}
               \hspace*{-0.01\textwidth} 
               \includegraphics[width=0.49\textwidth,clip=]{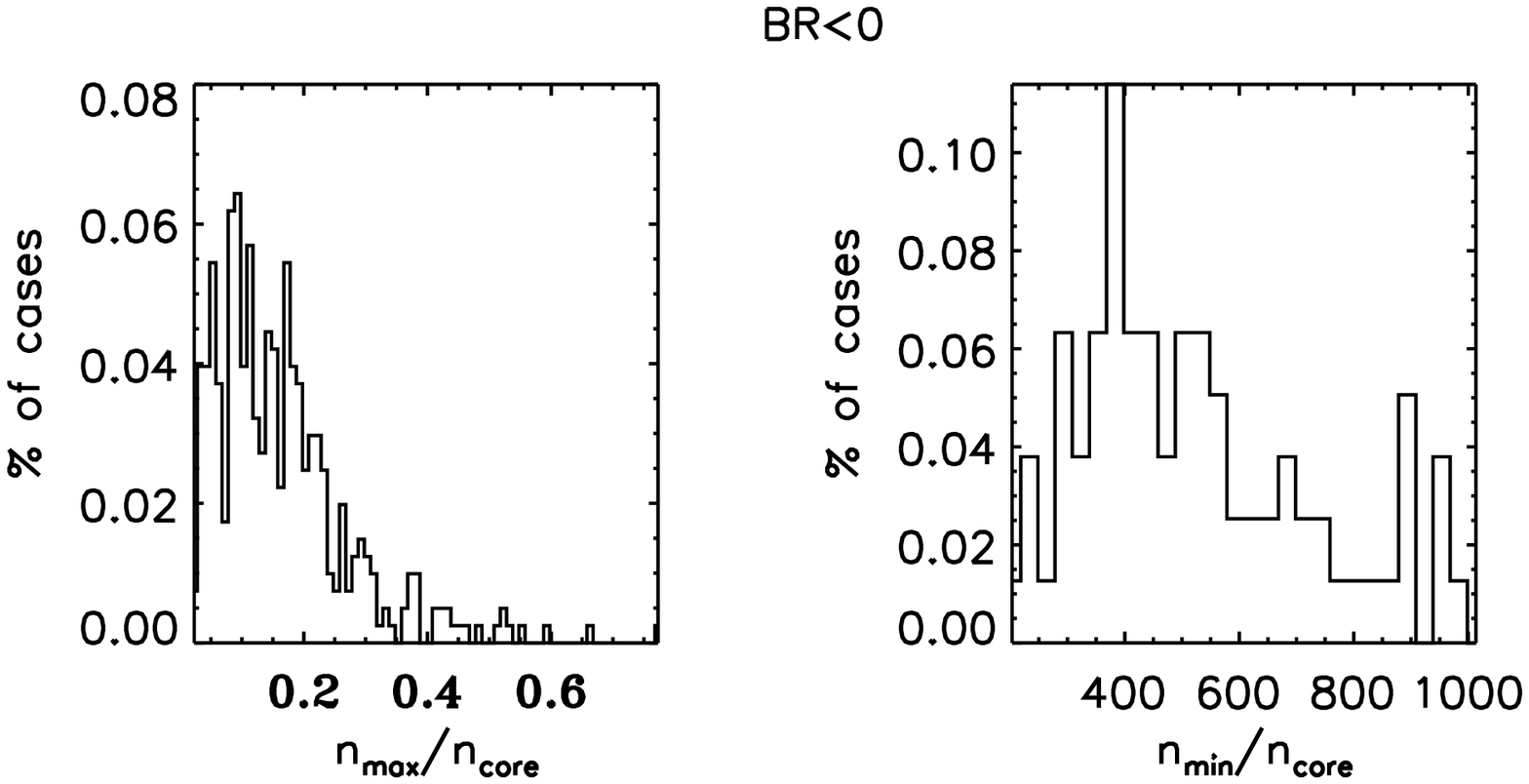}
              }
\caption{Histograms  $n_{max}/n_{core}$ and $n_{min}/n_{core}$
for the observation of 11 December
2007. Two left-most (right-most) panels correspond to cases with
positive (negative) blue-red asymmetry.} \label{fig:distr3}
\end{figure}

\begin{figure}
\epsscale{1.00}
\plotone{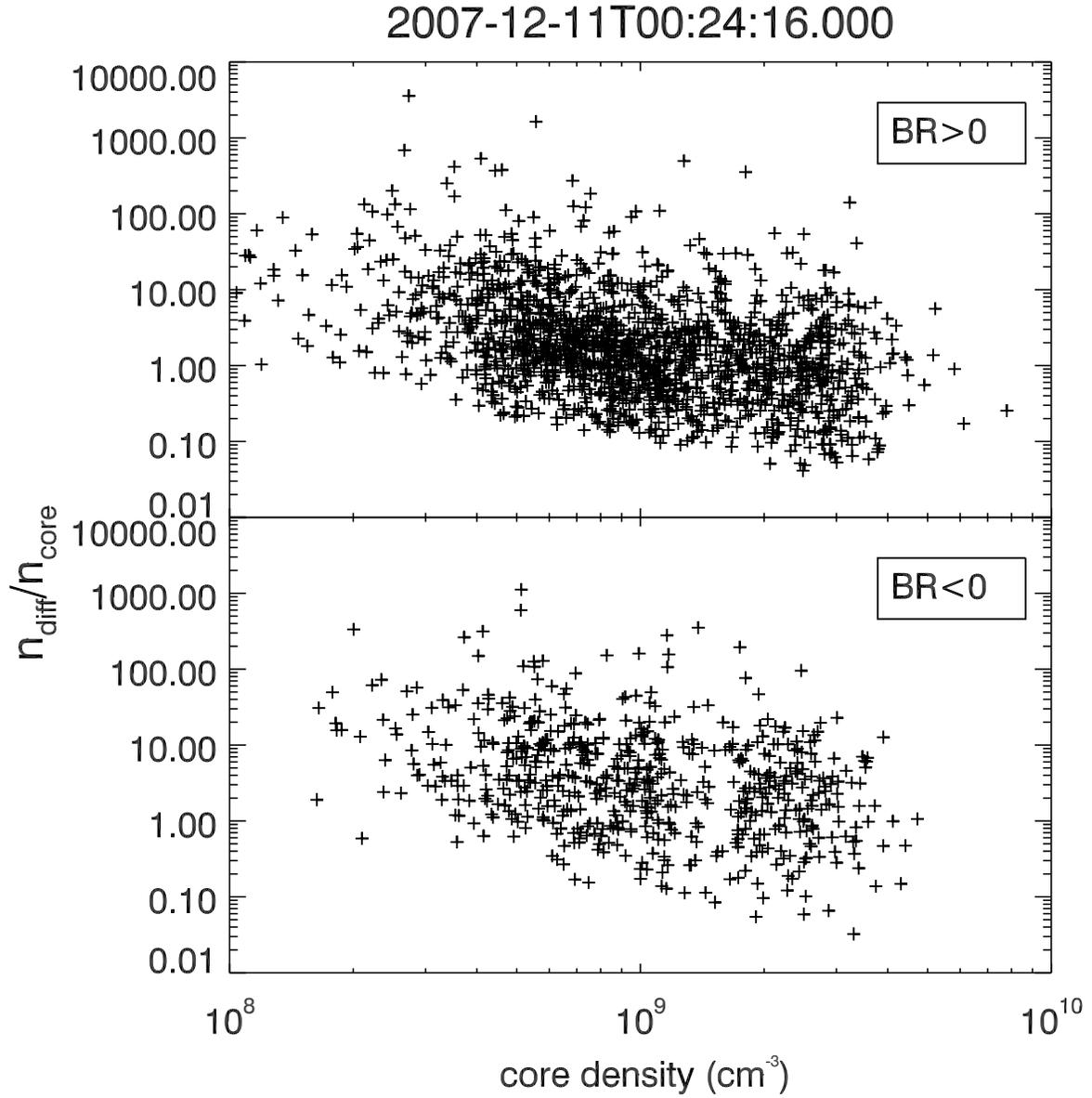}
\caption{$n_{diff}/n_{core}$ versus $n_{core}$ for profiles
exhibiting $BR>0$ (upper panel) and $BR<0$ (lower panel). Data
correspond to the base observation.}
\label{fig:fan}
\end{figure}

\begin{figure}
\epsscale{1.00}
\plotone{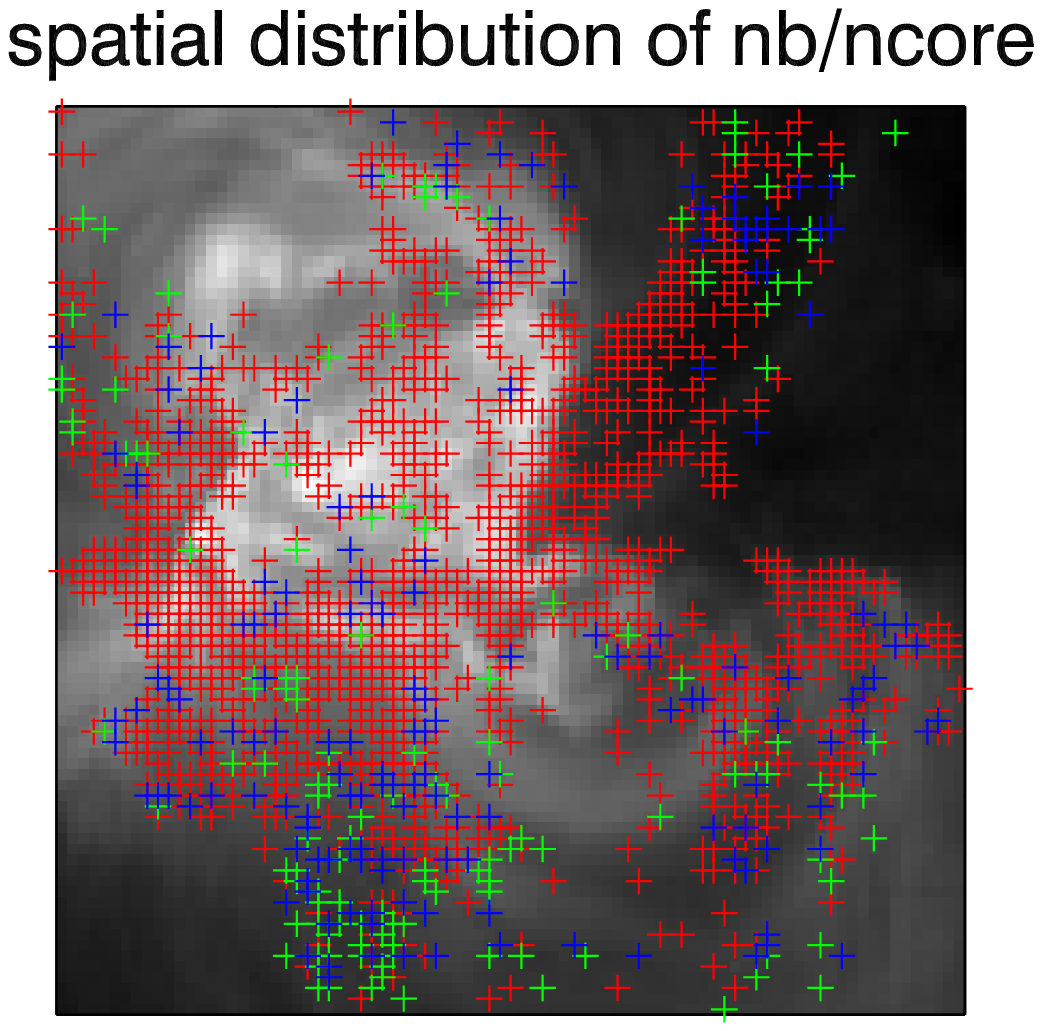}
\caption{Spatial distribution of
$n_{b}/n_{core}$ for the following ranges:
$\le 6.4$ (red crosses),
$\ge 16.6$ (green crosses)
$\in (6.4,16.6)$ (blue crosses), overplotted on
the corresponding 274 image.
Data
correspond to the base observation.}

\label{fig:spat}
\end{figure}

\begin{figure}
\epsscale{1.00} \plotone{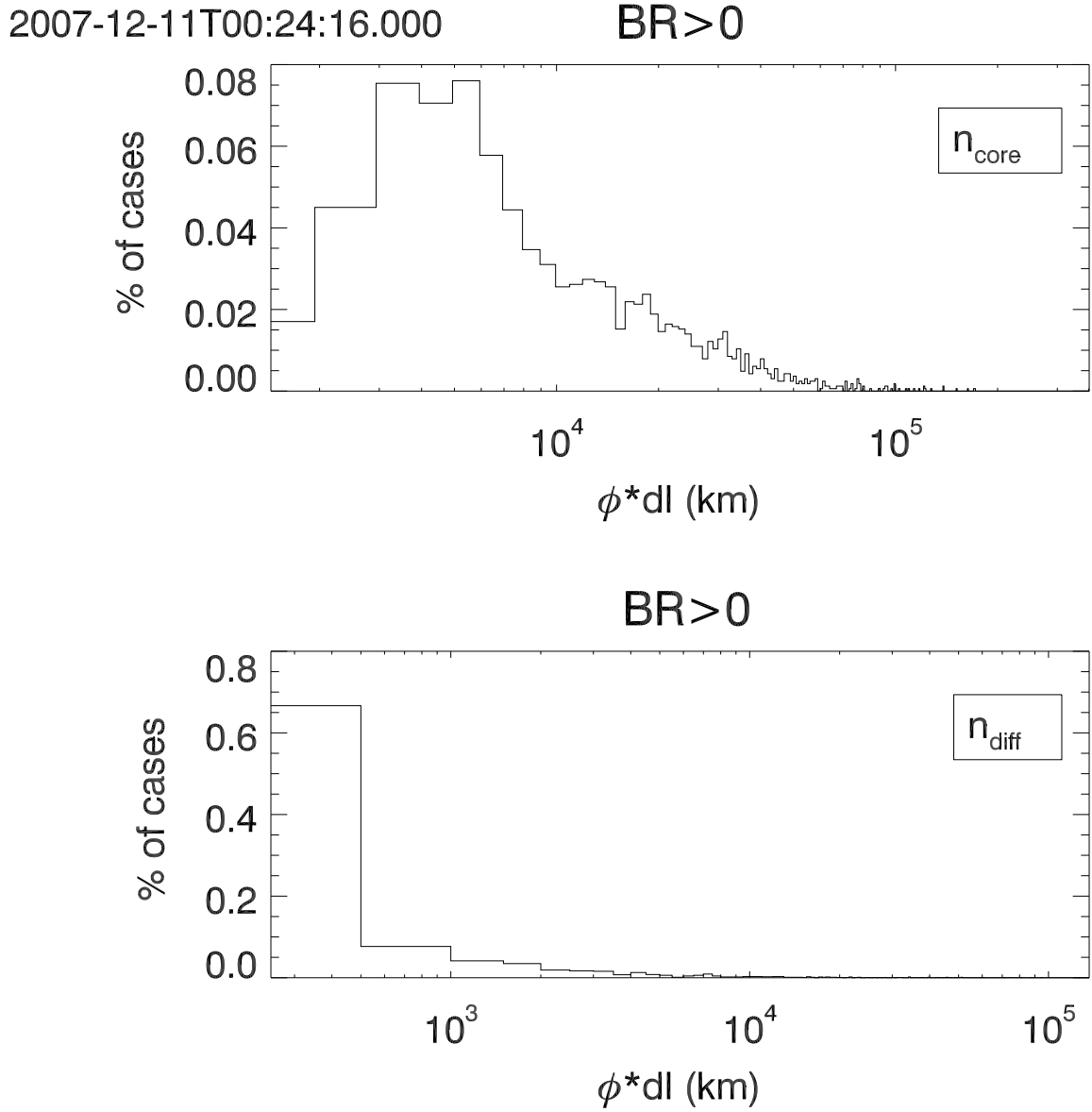} \caption{Histograms of
formations lengths for cases with $BR>0$ in the base observation.
Upper panel has the results for $n_{core}$ and lower panel for
$n_{diff}$.} \label{fig:ff}
\end{figure}

The main aim of this work is to establish the density
content of enhanced wing emissions far off
($\mid \delta v \mid \in [50, 150]$ \kms)
the line
centers and to compare it with that of the line cores
($\delta v \in [-30, 30]$ \kms), and particularly
for cases corresponding to blue-wing enhancements.
Our main conclusions
from the previous Sections are:
\begin{enumerate}
\item the bulk of the $n_{b}/n_{core}$ distribution,
where both quantities can be measured, (e.g., Figure
\ref{fig:distr2}) corresponds to too low values to be consistent
with the view that type II spicules are the primary source of hot
plasma in the corona.
\item the values are consistent with the predictions for coronal
nanoflares.
\item cases in the tail of the
 $n_{b}/n_{core}$ distribution
(e.g., Figure \ref{fig:distr2}) are
consistent with type II spicules
since they correspond to high
$n_{b}/n_{core}$ values.
The same
applies to cases with blue-wing enhancement density corresponding to the lower
density limit
(e.g., Figure \ref{fig:distr3}).
Both these cases correspond though
to a small fraction of macropixels
(for example the latter corresponds
to only the 2.9 $\%$ 
of macropixels
of the base observation).
\item
cases for which only an upper limit to the blue-wing enhancement density can be estimated
(e.g., Figure \ref{fig:distr3})
account for a significant fraction of the total number of macropixels 
(14.6 $\%$ 
for the base observation). These low density values are not consistent with type II spicules.
\end{enumerate}

\begin{figure}
\epsscale{1.00}
 \centerline{\hspace*{0.015\textwidth}
               \includegraphics[width=0.49\textwidth,clip=]{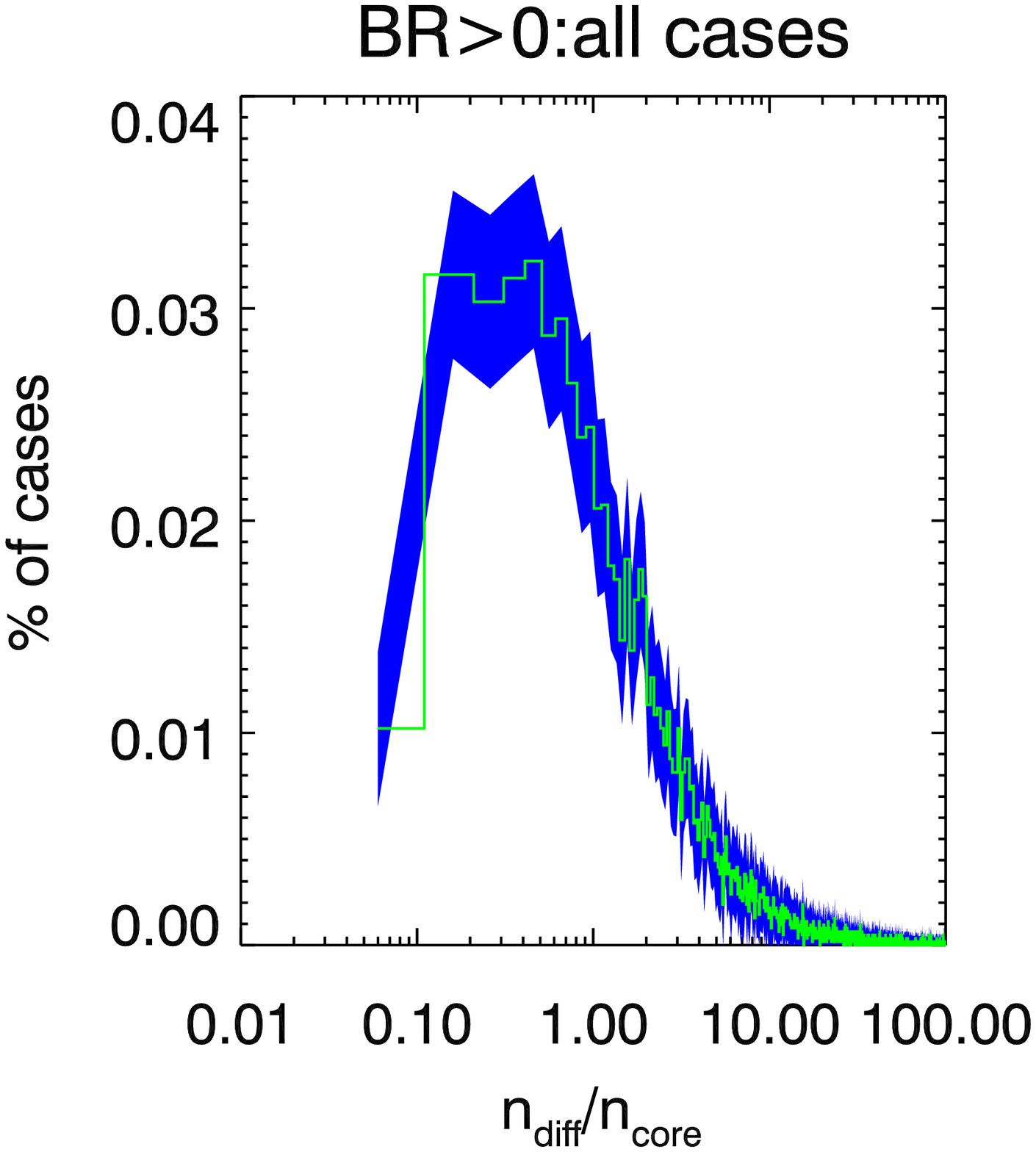}
               \hspace*{-0.01\textwidth}
               \includegraphics[width=0.49\textwidth,clip=]{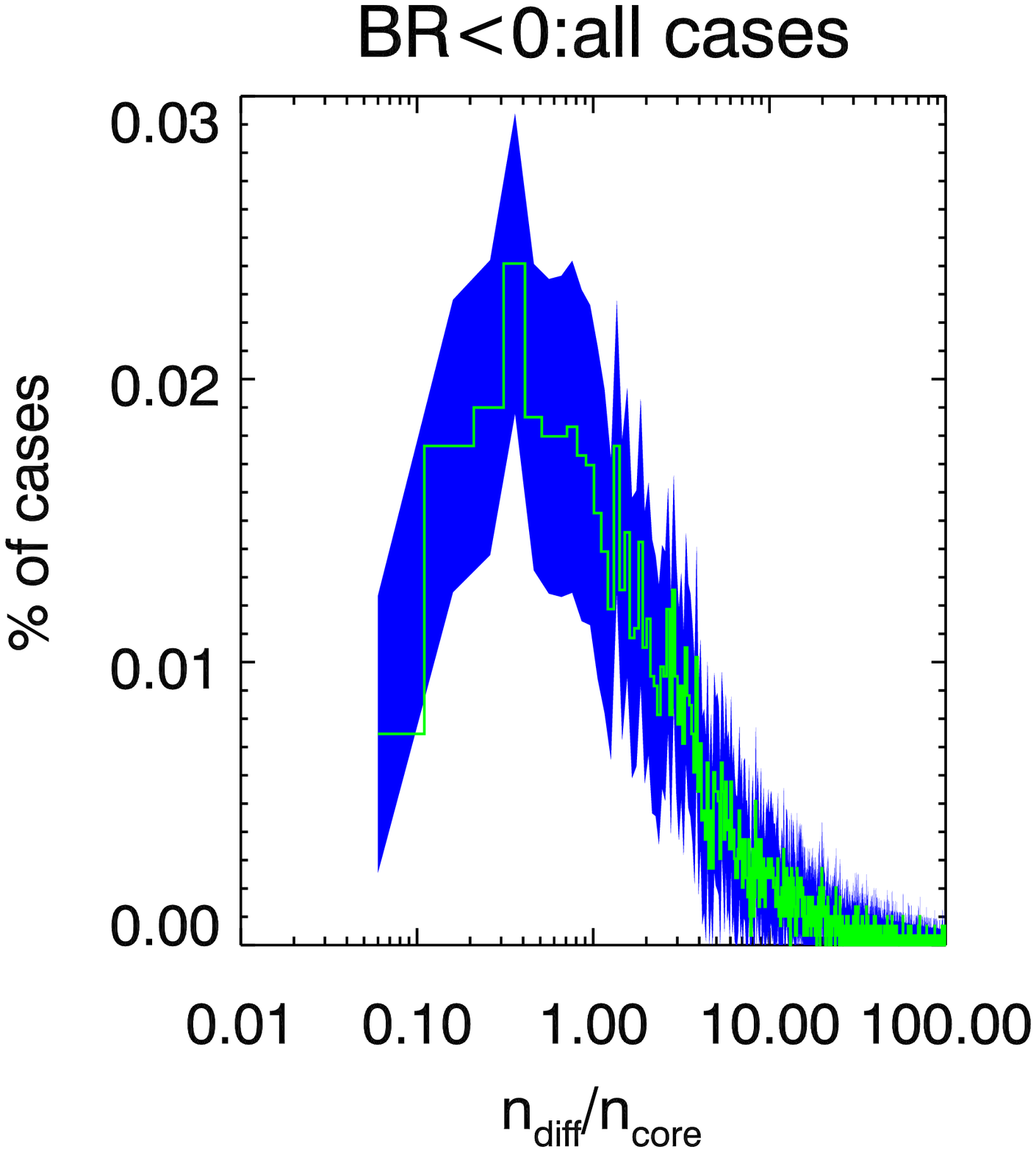}
              }
\caption{Histograms (green histogram-mode lines)
and 2-$\sigma$ uncertainties (blue contours)
of $n_{diff}/n_{core}$ for
the all analyzed datasets. Locations with positive blue-red
(left panel) and negative blue-red asymmetry (right panel).}
\label{fig:distall}
\end{figure}

Our conclusions above can be  extended to a larger statistical
sample containing 7 more AR datasets on-top of the base observation
(see Table-1, Table-2 and Table-3). Take for example Figure
\ref{fig:distall} where we show the $n_{b}/n_{core}$ histogram for
the 8 analyzed datasets. Once more we note that the bulk of the
$n_{b}/n_{core}$ distribution corresponds to small blue-wing
densities, with again a high $n_{b}/n_{core}$ tail. Similar results
to the base observation apply also to cases where the lower or upper
density limit of the blue-wing enhancement is reached. Indeed, if
spicules are dominant, then one would expect to see most of the wing
densities to be very high, i.e., reaching the lower density limit.
The result however (column I of Table 2) is that very few  pixels
(1.7-3.5 $\%$) actually reach very high densities. We therefore
conclude from the present analysis that type II spicules are
consistent with a only small fraction of the considered cases and
thus their potential for supply of the bulk of the coronal mass in
ARs seems rather limited. This is  consistent with two recent
studies. In \citet{brooks12} the emission measure of the fast
upflowing plasma was found to strongly peak at coronal temperatures
(1.2-1.4 MK), while it was almost an order of magnitude smaller at
transition region  temperatures (0.6 MK). This finding was
interpreted as the plasma producing the asymmetries having coronal
origin and possibly not being directly ejected from below and heated
on the fly. Brooks and Warren made their measurements in the faint
periphery of an active region, where the BR asymmetries are
greatest.  \citet{tri13} recently obtained similar results in the
bright core of an active region. Moreover, \citet{klim12} used
several independent arguments, including ones based on the line
asymmetry and on the ratio of emission measures in lower transition
region and corona, to conclude that type II spicules provide only a
small fraction ($<$ 2 $\%$) of the hot plasma observed in the AR
corona.

We now pass on to a discussion concerning cases with enhanced
red-wing emissions (i.e., $BR<0$). They represent a significant, yet
smaller, fraction compared to cases with $BR>0$ (for example they
correspond to 14.5 $\%$ of the total number of cases or about the
half of the cases with $BR>0$). The $n_{r}/n_{core}$, $n_r$ is the
red-wing enhancement density, distributions peak at around 1 for
both the base observations and the ensemble of the 8 AR
observations, right panels of Figures \ref{fig:distr2} and
\ref{fig:distall} respectively. There are only few cases at the
high-ratio tails of the $n_{r}/n_{core}$ distributions for which
$n_{r}$ is much larger than $n_{core}$. Finally, similar results to
the $BR>0$ case, are obtained for profiles with densities equal to
the upper and lower density limit (see for example panels 3 and 4 of
Figure \ref{fig:distr3}).

What could be then the physical mechanism causing these red-wing
enhancements? "Standard"  mass draining taking place during the
radiatively dominated cooling phase of cycles of impulsive heating
and cooling cannot possibly account for these enhancements. This is
because the predicted downflow speeds typically barely exceed
$\approx$ 20 \kms, which is much smaller than the speed of the
inferred downflows $>$50\kms. An attractive  possibility to generate
high-speed downflows can be catastrophic cooling and coronal rain
\cite[e.g.][]{anti99,karp01,mull04,karp08,mok08,anto10} The essence
of this mechanism is that highly concentrated heating at the feet of
coronal loops increases the coronal density until the coronal
radiative losses cannot any more be supported by (the weak) coronal
heating. This triggers a run-away catastrophic cooling and
high-speed downflows, and speeds as high as $\approx$ 100 \kms are
predicted. The observational signature of  catastrophic cooling is
thought to be the coronal rain, which consists of high-speed
downflowing blobs observed off-limb  \cite[e.g.][]{schri01}. Current
estimates of the fraction of the coronal volume occupied at any
given time by coronal rain are rather uncertain since they are
carried out in off-limb observations, with the inherent influence of
projection effects \citep{antooff}. On the other hand, observing
coronal rain on the disk is a more promising avenue for determining
the discussed above coronal rain filling factor, although plagued by
the low visibility of the associated blobs against the disk
background \citep{antodisk}. Coordinated disk observations of coronal
rain and of line asymmetries in coronal lines could help.

Our study, along with any other aiming at the wings of spectral
lines, is arguably pushing the analysis and the interpretation of
current spectroscopic observations to their limits. For example with
this study we took the utmost caution to minimize the impact of any
known blend for the two Fe XIV lines used in the
density determinations. This requirement already eliminates a
significant fraction of pixels from further analysis ($\approx$
11.1-58.2 \% of the pixels of the 8 analyzed datasets; see column B
of Table-2). Moreover, there is a significant fraction of pixels
with inconsistent sign of $BR$ asymmetry for the two Fe XIV lines
($\approx$ 18.4-35.9 \% of the pixels of the 8 analyzed datasets;
see columns E and F of Table-2). This implies the presence of some
identified blends at the line wings or maybe the impact of low S/N
at line wings. The main
reason that blends and/or noise force us to exclude many pixels from
our measurements is that the excess wing emission is extremely
faint. Klimchuk (2012) used this fact to conclude that type II
spicules are not a dominant source of hot coronal plasma. All the
above is essentially telling us that we cannot use measured
densities to comment on the source of the coronal mass for a
significant fraction ($\approx$ 42.0-78.3 \% of the pixels of the 8
analyzed datasets; see columns B,E,F in Table-2) of the observed
pixels. Moreover, the wavelength sampling of  EIS observations is
rather coarse when performing studies of the fine and subtle details
of spectral line profiles. A typical line profile (core and wings)
comprises no more than about 12 EIS spectral pixels which make it
difficult to resolve blends and to generally work on wing emissions.
Increasing the wavelength sampling and the sensitivity of
spectroscopic observations may help into increasing the number
observational pixels with useable wing emissions.

Another potential limitation of the present study is that our
density calculations are based on the assumption of ionization
equilibrium. Such an assumption is possibly suspect when dealing
with fast flowing plasmas associated with either type II spicules or
nanoflares given that they are formed under rapidly varying physical
conditions involving time-dependent heating and flow through steep
temperature gradients \cite[e.g.][]{brad06, real08, brad11}.
Detailed calculations of density-sensitive ratios of fast flowing
plasmas under dynamic conditions, taking into account possible
departures from ionization equilibrium, are required to determine
the precise impact of such effects on the corresponding diagnostics.
Indeed, this important task has been recently undertaken in
\citet{doyle12} and \citet{olluri13}, where density diagnostics for
the intensity ratio of two transition region lines of O IV was
calculated for dynamic events. The inclusion of non-equilibrium
ionization led to significant deviations of the resulting densities
(up to an order of magnitude underestimate) with respect to the
values calculated under ionization equilibrium conditions.

Next generation sub-arcsecond class spectrometers with
chromosphere-transition region-corona coverage like the Interface
Region Imaging Spectrometer (IRIS; \cite{iris}) the Very High
Angular Resolution Imaging Spectrometer (VERIS; \cite{veris}) and
the Large European Module for solar Ultraviolet Research (LEMUR;
\cite{ter12}) are expected to significantly advance our
understanding of the flow of the mass and heat in the solar
atmosphere. For example, their sub-arcsecond resolution would allow
to spectroscopically resolve type II spicules in the plane of the
sky, i.e., observe a distinct spectroscopic component and not merely
a weak wing enhancement. This separation would allow to take better
measurements of their physical parameters (densities, emission
measures, abundances, etc) and compare them with the plasma bulk.
However, there is always the inherent problem of LOS
integration, so the spicule emission will be mixed with all the
other emission in front of it.

\acknowledgments 
The authors thank the referee for useful and constructive comments
and suggestions which led to a significant improvement of the MS.
S.P. acknowledges support from FP7 Marie Curie
Re-integration Grant FP7-PEOPLE-2010-RG/268288 and from the European
Union (European Social Fund – ESF) and Greek national funds
through the Operational Program "Education and Lifelong Learning" of
the National Strategic Reference Framework (NSRF) - Research Funding
Program: Thales. Investing in knowledge society through the European
Social Fund.  The work of J.A.K. was supported by the NASA
Supporting Research and Technology Program.
The work of PRY was performed under contract with the Naval Research Laboratory and was funded by NASA.

\newpage

\begin{deluxetable}{clccc}
\tabletypesize{\scriptsize}
\tablecaption{General information of the analyzed datasets.}
\tablehead{
\colhead{Dataset} &
\colhead{Date} &
\colhead{Time (UT)} &
\colhead{Location (arcsec
from Sun center)} &
\colhead{Field of View Size}
}

\startdata
1 & 1  December 2006   &  00:24:16   & (-198,-141)  &  255$\times$255 \\
2 & 2 December 2006   &  10:33:23   & (203,-185)  &  255$\times$255 \\
3 & 20 Janurary 2007   &    02:33:52 & (347,9)  &  255$\times$255 \\
4 & 1 February 2007   &   00:12:12 & (-185,-141)  &  255$\times$255 \\
5 & 2 May 2007   & 05:27:17   & (122,-132)  &  240$\times$240 \\
6 & 1 July 2007   & 04:56:57  & (-112,-215)  &  240$\times$240 \\
7 & 24 August 2007   &  01:18:05  & (-755,-169)  &  255$\times$255 \\
8 & 11 December 2007   &  00:24:16   & (-198,-141)  &  255$\times$255
\enddata
\label{table:obs}
\end{deluxetable}

\clearpage

\begin{deluxetable}{ccccccccc}
\tabletypesize{\scriptsize}
\tablecaption{Macropixel  statistics. All columns except the first
supply percentages over the total number of macropixels in each
dataset.}
\tablewidth{0pt}
\tablehead{ \colhead{A\tablenotemark{1}} & \colhead{B\tablenotemark{2}} & \colhead{C\tablenotemark{3}}
& \colhead{D\tablenotemark{4}} & \colhead{E\tablenotemark{5}} & \colhead{F\tablenotemark{6}} & \colhead{G\tablenotemark{7}} &
\colhead{H\tablenotemark{8}} &
\colhead{I\tablenotemark{9}} \\
\hline
\colhead{Dataset} & \colhead{``No'' Blend} & \colhead{B \&} &
\colhead{B\&} &
\colhead{B \& } &  \colhead{B \& } & \colhead{C(D) \&  } &
\colhead{C(D) \& } & \colhead{C(D) \&} \\
\colhead{Number} & \colhead{} & \colhead{$I_{B}>I_{R}$} &
\colhead{$I_{B}<I_{R}$} &
\colhead{$I_{B}>I_{R}$} &  \colhead{$I_{B}>I_{R}$} & \colhead{measurable $n_{diff}$ } &
\colhead{high $n_{diff}$} & \colhead{low $n_{diff}$} \\
\colhead{} & \colhead{} & \colhead{both lines} &
\colhead{both lines} &
\colhead{264 only} &  \colhead{274 only} & \colhead{ } &
\colhead{  } & \colhead{ } \\
\colhead{} & \colhead{} & \colhead{} &
\colhead{} &
\colhead{} &  \colhead{} & \colhead{ } &
\colhead{limit} & \colhead{limit} \\
}
\startdata

1 & 52.3    &  21.2    & 11.7     &  6.1  &  13.2   & 8.9(5.1)  & 10.3(3.9)  & 1.7(2.6)  \\
2 & 72.3   &  28.2    & 14.5      &  7.2  &  22.3   & 12.3(6.8)  & 13.2(4.7)  & 2.5(2.9)  \\
3 & 58.6   &  29.6    & 8.8     &  7.0  &  12.9     & 13.1(3.9)  & 13.3(2.4)  & 3.1(2.4)  \\
4 & 75.9   &  30.4    &  13.6      &  7.7  &  24.0  & 11.5(6.6)  & 15.2(3.1)  & 2.5(3.5)  \\
5 & 65.0    &  22.0    & 13.0      &  7.5  &  22.3  & 8.6(6.0)  & 10.2(3.5)  & 2.5(3.2)  \\
6 & 41.8    &  12.8    & 8.8     &  7.1  &  13.0    & 5.0(3.2)  & 4.2(2.3)  & 2.4(2.8)  \\
7 & 88.9    &  34.6    & 18.3      &  11.5  &  24.4  & 13.7(7.5)  & 12.3(5.0)  & 3.5(4.1)  \\
8 & 76.4    &  41.3   & 16.6      &  9.1  &  9.3  & 22.7(8.4)  & 14.6(5.5)  & 2.9(2.5)
\enddata
\tablenotetext{1}{Dataset number.}
\tablenotetext{2}{Exhibiting
``no blend'' (i.e., maximum Si VII and Fe XI intensities  $<10\%$ of
the 264 and 274 Fe XIV intensities, respectively (see the discussion
in Section \ref{sec:anal}).}
\tablenotetext{3}{Condition of column B and
$I_{B}>I_{R}$ for both 264 and 274.}
\tablenotetext{4}{Condition of
column B and $I_{B}<I_{R}$ for both 264 and 274.}
\tablenotetext{5}{Condition of
column B and $I_{B}<I_{R}$ for both 264 and 274.}
\tablenotetext{6}{Condition of column B and $I_{B}>I_{R}$ for 274 only.}
\tablenotetext{7}{Condition of columns C or D (quantities in parentheses) and
$n_{diff}$ in the density sensitive part of the $n-I_{264}/I_{274}$
curve (see Section \ref{sec:diagn}).}
\tablenotetext{8}{Condition of
column C or D (quantities in parentheses) and $n_{diff} < $ upper
density limit ($\equiv {10}^{8} \mathrm{{cm}^{-3}}$).}
\tablenotetext{9}{Condition of column C or D (quantities in parentheses) and $n_{diff}
> $ lower density limit ($\equiv {10}^{12} \mathrm{{cm}^{-3}}$).}
\label{table:gstats}
\end{deluxetable}

\clearpage

\begin{deluxetable}{cccc}
\tabletypesize{\scriptsize}
%\rotate
\tablecaption{Statistics of $n_{diff}/n_{core}$ for cases
corresponding to the entries of column G of Table 2,
i.e. macropixels
where $n_{diff}$ and $n_{core}$ can be measured
and BR is positive or negative (in parentheses) for both 264 and 274.}
\tablewidth{0pt}
\tablehead{
\colhead{Dataset Number} &
\colhead{Peak of distribution} &
\colhead{Median of distribution} &
\colhead{$FWHM$ of distribution}
}
\startdata
1 & 0.4(1.0)   &  2.69(5.3)  & 0.9 (0.9)   \\
2 & 0.6(2.8)   &  2.2(4.4)  &  1.3(2.3)   \\
3 & 0.7(0.1)   & 2.8(5.4)   & 0.9(1.7)    \\
4 & 0.4(0.3)   & 3.9(4.3)   & 1.3(1.5)   \\
5 & 0.8(0.2)   & 3.3(4.3)   & 1.3(0.8)    \\
6 & 0.1(1.1)   &  3.0(8.9) &  0.4(1.2)   \\
7 & 0.5(0.3)   &  4.9(6.8)  & 1.3(0.4)   \\
8 & 0.1(0.3)   & 1.6(3.3)   & 1.0(1.2)
\enddata
\label{table:diststats}
\end{deluxetable}

\clearpage

\clearpage

\clearpage

\clearpage

\clearpage

\clearpage

\clearpage

\clearpage

\end{document}